\documentclass[twocolumn,showpacs,aps,prl,nofootinbib,superscriptaddress]{revtex4-1}
\usepackage{amsmath,graphicx,amssymb,slashed}
\newcommand{\comment}[1]{}

\def\Pmatrix#1{\begin{pmatrix} #1 \end{pmatrix}}
\def\half{ {1\over 2}}

\begin{document}
\begin{flushright}
SLAC-PUB-17255
\end{flushright}

\title{Why the angular distribution of the top decay lepton is
     unchanged by  anomalous $tbW$ couplings}
\author{Rohini M. Godbole}
\email{rohini@cts.iisc.ernet.in}
\affiliation{Center for High Energy Physics, Indian Institute of Science,
Bangalore  560012, India}
\affiliation{Institute of Physics and Astronomy, University of Amsterdam, Science Park 904 , The Netherlands}
\author{Michael E. Peskin}
\email{mpeskin@slac.stanford.edu}
\affiliation{SLAC, Stanford University, Menlo Park, CA 94025, USA}
\author{Saurabh D. Rindani}
\email{saurabh@prl.res.in} 
\affiliation{Theoretical Physics Division, Physical Research Laboratory,
Navrangpura, Ahmedabad  380009, India }
\author{Ritesh K. Singh}
\email{ritesh.singh@iiserkol.ac.in} 
\affiliation{Department of Physical Sciences,
Indian Institute of Science Education and Research Kolkata,
Mohanpur  741246, India }

\begin{abstract} 
We give a simple physical argument to understand the observation that the 
angular distribution of the top decay lepton depends only on the polarisation 
of the top and is independent of any anomalous $tbW$ coupling to linear order. 
\end{abstract} 
\maketitle
\def\bea{\begin{eqnarray}}
\def\eea{\end{eqnarray}}
\def \f{\frac}
The top quark is the heaviest known fundamental particle. Its
average lifetime is about one order of magnitude smaller than the typical
hadronisation time scale. This leads to decay of the top quark
before the strong-interaction
hadronisation process can wipe out its spin information. Thus, one can 
extract the top quark polarisation from the kinematical distributions
of 
  its decay products. 

The polarisation of the $t$ quark produced via Standard Model (SM) processes at hadron
colliders
 is known.
It is zero for the dominant QCD-induced $t \bar t$ production and is
dominantly left-handed but 
calculable for the subdominant single-$t$ production.  The rigidity of
these  predictions allows us to use the $t$ polarisation to probe for
 possible new physics contributions to these production processes.
 From simple angular momentum considerations, the angular distribution
 of a  spin $1/2$
decay product $f$ of the $t$ quark must take the form 
\begin{equation}
\frac{1}{\Gamma_t} \frac{d\Gamma_t}{d\cos\theta_f} = \frac{1}{2} \big(
1+ \alpha_f P_t \ \cos\theta_f\big)
\label{theta}
\end{equation}
In the SM, one finds for the $t\to b \ell \nu $ decay the values 
$\alpha_b=-0.4$, $\alpha_{\ell}=1$ and $\alpha_{\nu}=-0.32$ 
at tree level, and only small modifications of these values at the
one-loop level.   Using these values, the  measurement of the $t$
decay angular distributions 
can be used to obtain the $t$ polarisation.   

However, there is a possible problem.  If new physics can modify the
$t$ quark production amplitudes, it can also modify the $t$ quark
decay amplitudes.   Then we would expect new physics to modify the
values of the $\alpha_i$ in Eq.~(\ref{theta}).    The measurement of
the distributions gives only the combinations $\alpha_i P_t$, so if
the $\alpha_i$ can be shifted by new physics effects, this method loses its power.

Thus it is noteworthy that, 
in a series of investigations on $\bar tt$ production at an 
$e^+e^-$ collider~\cite{decouplingee,sdrPramana,decouplingee2}
and a $\gamma\gamma$ collider~\cite{decouplinggg, Godbole:2002qu},
it  was observed that $\alpha_\ell$ remains
unchanged even after inclusion of anomalous $tbW$ couplings, up to
linear order in new physics parameters. This independence of
$\alpha_\ell$
 (or ``lepton decoupling") was also observed for more 
 general processes  of top-quark production 
\cite{Godbole:2006tq,Godbole:2010kr}, suggesting that it is 
 a property of the top quark decay and not of any specific production process. 
This would make the angular distribution of the decay lepton with
respect to the top spin direction a very robust measure of the top
polarisation.

It was also
 noted \cite{Godbole:2006tq,Godbole:2010kr} that this lepton
 decoupling follows
 because, for the SM, the full kinematic
 distribution of the decay lepton 
 factorises into a term  dependent on the
 lepton energy $E_{\ell}$ and another term dependent
 on the angular variables, and that this factorisation
 is maintained even in the presence of anomalous
 $tbW$ couplings up to linear order.   Actually, the dependence of the
 decay distributions on $E_\ell$ is modified by anomalous $tbW$
 couplings. Then it is possible to  use the angular and energy
 distributions
 together to measure the polarisation of
 the $t$ quark and in addition to probe for the presence of anomalous
 couplings in the decay vertex~\cite{Godbole:2006tq,agr,as}. 

Both the lepton decoupling and 
the factorization do receive corrections at the
 quadratic order in anomalous
 couplings~\cite{Godbole:2002qu,AguilarSaavedra:2006fy}. 
But, in view of the already rather strong constraints on 
 the $tbW$ vertex~\cite{PDG}, in which the least constrainted parameter
 $f_{2R}$ is required to be less than about 0.1,  lepton decoupling to
 linear order is quite sufficient for practical purposes. 

In Ref.~\cite{Hioki:2015moa}, Hioki has given an argument for lepton
decoupling based on a physical picture.  In this paper, we would like
to present a more transparent derivation of this result.

\textit{Derivation:}
The key ingredient in our proof of lepton decoupling 
 is the fact that, in the SM, the $b
\nu_{\ell}$ system produced in $t \rightarrow b \ell \nu$ is in
a $J=0$ state. As a result of this,  the entire spin of the
top is transferred to the lepton. This can be seen by a 
Fierz transformation of the SM decay
amplitude. Starting from this fact, we will show that lepton
decoupling for anomalous terms in the $tbW$ vertex follows from 
simple rotation algebra.  

At the tree level in the SM, the amplitude for $t$ decay is a product
of matrix elements of left-handed currents.  Using
$$   P_L = {1-\gamma^5\over 2 } = \Pmatrix{1 & 0\cr 0& 0\cr }\qquad
\gamma^\mu = \Pmatrix{ 0 & \sigma^\mu \cr \bar\sigma^\mu & 0\cr}  $$
and considering only the upper two components of the Dirac spinors, we
can write the decay matrix element as
\begin{equation}
 i{\cal M}  = i G(p_W) \ 
      u^\dagger(b) \bar\sigma^\mu u(t) \   u^\dagger(\nu)
      \bar\sigma_\mu v(\ell) \ ,
\label{myM}
\end{equation}
with
$$  G(p_W) =  {g^2\over 2} V_{tb}  {1\over p_W^2 - m_W^2}  \ .
$$
Then the Fierz identity
$$   ({\bar\sigma}^\mu)_{ab} \ (\bar\sigma_\mu)_{cd} = 2 \epsilon_{ac} \epsilon_{bd} 
$$ 
converts Eq.~(\ref{myM}) into 
\begin{equation} i{\cal M}  =   2 i\  G(p_W) \ 
   [   u_{a}^\dagger(b) \epsilon_{ab} u_{b}^\dagger(\nu) ]\ [u_{c}(t) \epsilon_{cd} v_{d}(\ell) ] \ .
\label{myMtwo}
\end{equation}
Each bracket is a Lorentz-invariant.   In particular, the $(b\nu)$
system is produced in a $J=0$ state.

We can now use the result in Eq.~(\ref{myMtwo}) to compute the spin
density matrix for the $t$ quark in terms of the lepton orientation.
We will do this first in the SM and then add anomalous $tbW$ couplings
to linear order.

\begin{figure}[ht]
\includegraphics[width=6.6cm]{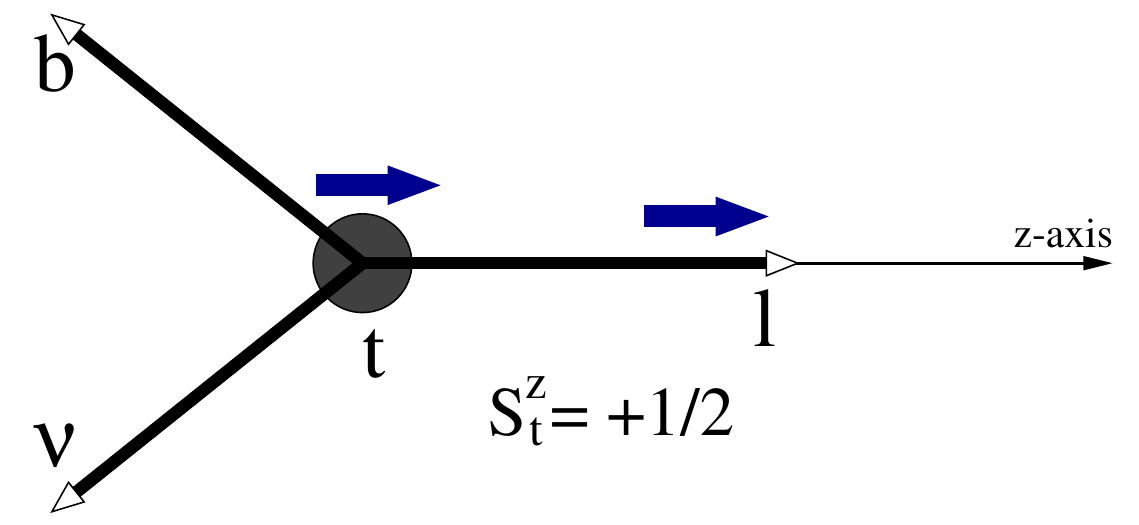}\hfill
\includegraphics[width=6.6cm]{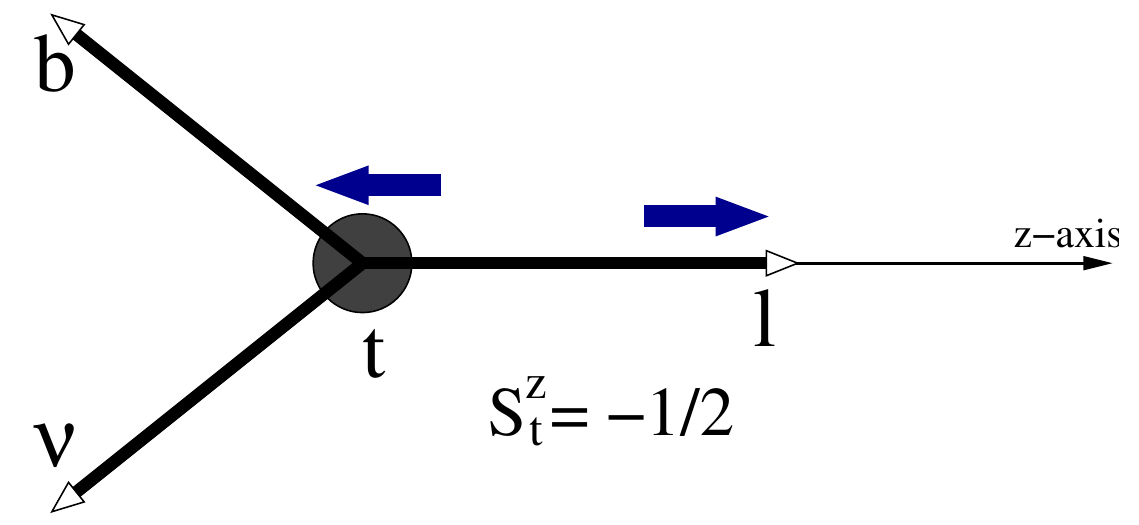}
\caption{\label{fig:conf}The configurations with the lepton momentum
along the top spin
quantization axis, the  $z$ axis.}
\end{figure}

The  decay lepton produced in (an assumed SM)  $W$ decay is
always right-handed. Hence the lepton direction is correlated with the lepton spin.
We work in the rest frame of the decaying $t$ quark.  The $t$ spin
orientation is defined by a 2-component spinor $\xi$ in the frame of the
decay.  Then we can best analyze the density matrix by choosing coordinates
in which the lepton momentum is parallel to the $\hat z$ axis.  
The decay amplitude is a linear combination of the amplitudes for two
configurations, those in which the $t$ spin is parallel and  antiparallel to
the $\hat z$ axis.  We show these two cases in Fig.~\ref{fig:conf}.

In the SM, it follows directly from Eq.~(\ref{myMtwo}) that the decay
amplitude for $t$ spin $S_t^z = -\half$ vanishes. Then the spin
density matrix takes the form
$$  \Gamma_t =   F(E_\ell)
\Pmatrix{1& 0\cr 0 & 0 \cr} \ , 
$$
 Already here we
see the factorization of the dependence on $t$ spin and lepton energy.
To obtain the density matrix for a general $t$ spin orientation---or
for a general orientation  $(\theta_\ell, \phi_\ell)$ of the lepton
direction 
relative to the 
$\hat z$ axis---we perform a rotation and obtain
\begin{equation} \Gamma_t =  F(E_\ell) \
\Pmatrix{(1 + \cos\theta_\ell) &\sin \theta_\ell e^{i\phi_\ell} \cr
 \sin \theta_\ell e^{-i\phi_\ell}  &(1 - \cos\theta_\ell) \cr} \ .
\label{myfirstG}
\end{equation}
in accord with \cite{decouplingee2}.

Now introduce a general set of form factors representing anomalous
couplings in the $tbW$ vertex,
\begin{eqnarray}
\Gamma^\mu&=&\frac{-i g}{\sqrt{2}} \ \left[\gamma^\mu((1+f_{1L})P_L + f_{1R}P_R)
\right. \nonumber\\ &-& \left. 
\frac{i\sigma^{\mu\nu}}{m_W} \ (p_t-p_b)_\nu \ (f_{2L}P_L +
f_{2R}P_R) \right.]
\label{V:tbW}
\end{eqnarray}
The SM vertex is the case in which  all coefficients $f_i$  are zero.

Analyze this more general situation in the frame shown in
Fig.~\ref{fig:conf}.   Now the decay matrix elements for $S_t^z = +\half$
and $S^z_t = -\half$ are both nonzero, and thus
 all four elements of the $t$ spin density
matrix receive nonzero contributions either linear or quadratic  in the $f_i$.  However,
the new contributions to the decay amplitudes can depend on the angle
between the plane containing the $(b,\nu)$ momentum vectors and the
reference $(\hat x, \hat z)$ plane.  Call this angle $\phi_b$.  For
$S_t^z = +\half$, the $(b,\nu)$ system has $S^z = 0$ and so the decay
amplitude is independent of $\phi_b$.  On the other hand, for $S^z_t
= -\half$, the $(b,\nu)$ system must carry away $S^z = -1$, and so 
$$  i{\cal M} (\phi_b) = i {\cal M}(\phi_b = 0) \cdot
e^{i\phi_b}\  .  $$
The density matrix then takes the form
$$
\Gamma_t \propto 
\Pmatrix{ 
1+{\cal O}(f_i) & {\cal O}(f_i)\cdot e^{-i\phi_b}\cr 
 {\cal O}(f_i)\cdot e^{+i\phi_b} &  {\cal O}(f_i^2) \cr}
\label{eq:Gconf}
$$

To obtain the density matrix for the charged leptons, 
we integrate over  the orientations of the other $t$ decay products,
keeping the lepton momentum fixed.
This includes an integration over $\phi_b$.
After this integration, we find 
$$
\Gamma_t \propto
\Pmatrix{ 
1+{\cal O}(f_i) & 0\cr 
0 & 0 \cr}
\label{eq:Gconftwo}
$$
up to terms of quadratic order in the $f_i$.   Notice that the upper
left matrix element of $\Gamma_t$ can be modified by nonzero $f_i$,  in a manner that
depends on $E_\ell$; however, the factorization between the
dependences on $t$ spin and $E_\ell$ is preserved.  This is the result that
we sought to prove. In this case,  the density matrix for a general $t$ spin orientation 
will be similar to Eq.~\ref{myfirstG}, but with a different $E_\ell$ dependent factor.

It is useful to take stock of what we needed to assume, and what we
did not need  to assume, to 
achieve this result:
\textit{(i) Chiral lepton:}  We treated the $\ell^+$as  massless, and,
in accordance with the 
$V-A$ nature of the  $W$ boson decay, having strictly
positive helicity. 
\textit{(ii) SM spin correlation:} We needed the property of the
SM amplitude that the $t$ spin is completely correlated with the
lepton spin.   This property does not hold if 
we replace the charged lepton with either $\nu$ or $b$. 
So the result holds only for charged leptons,  and
 for $T_3=-1/2$ light quarks in hadronic decays of $W^+$-boson.
\textit{(iii) Partial averaging:} We needed to average over the
azimuthal orientation of the $b, \nu$ vectors in the frame of the $t$
decay.  This would naturally be done if the $t$ polarisation is
measured from the inclusive lepton distribution.
 
On the other hand, we did not require the $b$ quark to be massless
 or the $W$ boson to be
on-shell.   For 
massive leptons, viz., $\tau$'s, the matrix element 
${\cal M}(t_{\downarrow}\to \tau^+_{\uparrow} \ b\nu_\tau;\phi_b)$ does not 
vanish in the SM and we get $\alpha_\tau \neq 1$. This leads to a correction
in $\alpha_\tau$ at ${\cal O}(f_i)$, but this correction is suppressed by $m_\tau/m_t$.\footnote{We argue in this paper that the decay lepton distribution is a robust measure of the top quark polarization given by the production process, even in the case where there is an anomalous tbW coupling.  In single-top production, the anomalous tbW coupling contributes to the production cross section and so an anomalous contribution affects the top quark polarization that is
generated~\cite{Rindani:2011pk,Rindani:2011gt,Aguilar-Saavedra:2014eqa,Aguilar-Saavedra:2017nik,Jueid:2018wnj}.   
This does not affect our conclusions; the altered top quark polarization is still measured correctly by the lepton distribution.}

\textit{Conclusions:}
In this note, we have analyzed the robustness
 of the parameter $\alpha_{\ell}$ associated with the $t$ spin
 polarisation against the contributions from anomalous
$tbW$ couplings. We related this robustness to the factorisation of 
the energy and angle distributions for charged leptons. 
This factorisation emerges due to the SM property of the  vanishing of the 
amplitude for the charged lepton with momentum in a direction 
opposite to the top spin.   Further, the factorisation, demonstrated here in
the rest frame of the decaying quark, remains true in the laboratory frame as well. 
Thus energy integrated angular distribution of the lepton produced  in the decay of a
polarised top quark  does not receive any modifications from the anomalous $tbW$
 coupling,  in the laboratory frame  as well.

This analysis offers us insight into the effect of anomalous $tbW$
couplings on the kinematic distributions of the charged lepton
produced in the $t$ decay. The same analysis applies, in fully
hadronic $W$ decays, to the angular and energy distribution of the 
$T^3 = -\half $ quark in the final state~\cite{Ester,Tweedie}.  
The robustness of the independence of the
angular distribution from the anomalous couplings,  to linear order,
offers us the possibility of using these kinematic distributions
to construct independent probes of both the top
 polarisation and the anomalous $tbW$ couplings.

\vspace{0.5cm}
\noindent \textbf{Acknowledgments:} 
We thank Xerxes Tata for a careful reading and tough critique of this paper.
We are grateful to Stefano Frixione  and the CERN Theory Group for providing a
congenial atmosphere to begin our discussions.  MEP is grateful to   the Center for High Energy Physics at the Indian Institute of
Science, Banagalore, for a very pleasant setting in which to complete
them. We also thank Eric Laenen for his comments 
on the manuscript. The work of RMG is supported by the Department of Science and Technology, India 
under Grant No. SR/S2/JCB-64/2007.  SDR acknowledges support from the Department of Science and
Technology, India, under the J.C. Bose National Fellowship program,
Grant No. SR/SB/JCB-42/2009.   The work of MEP is supported by the 
US Department of Energy, contract DE--AC02--76SF00515. RMG wishes to acknowledge the support 
of the Institute of Physics and  Astronomy, Amsterdam for the  IPA Visiting Professorship in 2018 and the 
hospitality of the theory group at NIKHEF, Amsterdam.



\end{document}